\documentclass[12pt]{article}
\usepackage{a4wide,amssymb,cite}
\pdfoutput=1

\usepackage{a4wide,amssymb,graphicx}
\usepackage{epsfig}
\usepackage[usenames,dvipsnames]{color}
\usepackage{slashed}
\usepackage{amssymb,cite,graphicx}
\usepackage{slashed}
\usepackage{amsmath,bm,bbm}
\usepackage{amsfonts}
\usepackage[titletoc,title]{appendix}
\usepackage[small]{caption}
\usepackage[margin=1in]{geometry}
\usepackage[multiple]{footmisc}
\usepackage{mathtools}
\usepackage{slashed}
\usepackage[nottoc]{tocbibind}
\usepackage{xcolor}

\usepackage{tikz}
\usetikzlibrary{shapes,arrows,positioning,automata,backgrounds,calc,er,patterns}
\usepackage[compat=1.1.0]{tikz-feynman}
\usetikzlibrary{trees}
\usetikzlibrary{decorations.pathmorphing}
\usetikzlibrary{decorations.markings}

\newcommand{\be}{\begin{equation}}
\newcommand{\ee}{\end{equation}}
\newcommand{\bea}{\begin{eqnarray}}
\newcommand{\eea}{\end{eqnarray}}

\def\circa#1{\,\raise.3ex\hbox{$#1$\kern-.75em\lower1ex\hbox{$\sim$}}\,}

\begin{document}

\begin{titlepage}
%
%


%

\begin{centering}
\vspace{1cm}
{\Large {\bf Seesaw lepton masses and muon $g-2$  \vspace{0.2cm}\\ from heavy vector-like leptons}} \\

\vspace{1.5cm}

{\bf Hyun Min Lee$^{*,1}$, Jiseon Song$^{\dagger,1}$  and Kimiko Yamashita$^{\star,1,2}$ }
\vspace{.5cm}

{\it  $^1$Department of Physics, Chung-Ang University, Seoul 06974, Korea.} 
\\[0.2cm]
    {\it $^2$Institute of High Energy Physics, Chinese Academy of Sciences, Beijing 100049, China.}

\vspace{.5cm}


\end{centering}
\vspace{2cm}

\begin{abstract}
\noindent
We propose a model for the vector-like lepton to explain the small muon mass by a seesaw mechanism, based on lepton-specific two Higgs doublet models with a local $U(1)'$ symmetry. There is no bare muon mass for a nonzero $U(1)'$ charge of the leptophilic Higgs doublet, so the physical muon mass is generated due to the mixing between the vector-like lepton and the muon after the leptophilic Higgs doublet and the dark Higgs get VEVs. In this scenario, the non-decoupling effects of the vector-like lepton give rise to leading contributions to the muon $g-2$, thanks to the light $Z'$ and the light dark Higgs boson. We discuss various constraints on the model from lepton flavor violation, electroweak precision and Higgs data, as well as collider searches.

\end{abstract}

\vspace{3cm}

\begin{flushleft} 
$^*$Email: hminlee@cau.ac.kr \\
$^\dagger$Email: jiseon734@gmail.com  \\
$^\star$Email: kimiko@ihep.ac.cn 
\end{flushleft}

\end{titlepage}

\section{Introduction}

There has been a lot of renewed interest in new physics scenarios for the anomalous magnetic dipole moment of the muon, after the Fermilab E989 experiment \cite{fermilab} confirmed the previous results of the muon $g-2$ at Brookhaven E821 experiment \cite{amu-exp}.  The deviation of the experimental value of the muon $g-2$ from the theoretical prediction in the Standard Model \cite{amu} is now at $4.2\sigma$. There is a need of improvement for the estimation of hadronic contributions to the muon $g-2$ in the Standard Model (SM) \cite{epem}, and the recent lattice-QCD calculation with reduced errors has hinted at a smaller deviation \cite{bmw}, although the inferred $e^+e^-$ hadronic cross section below about $1\,{\rm GeV}$ is in tension with the experimental data \cite{passera,epemdata} and the global electroweak fit \cite{passera1,passera,ewfit}.
Nonetheless, it would be worthwhile to develop new physics ideas for the muon $g-2$ and test them by ongoing and future experiments. In order to explain the muon $g-2$ anomaly of about $\Delta a_\mu\sim 2.5\times 10^{-9}$, either heavy charged particles with a large violation of chirality or neutral light particles with feeble couplings are favored due to the null discovery of new charged particles at the weak scale.

We consider a model for the $SU(2)_L$ singlet vector-like lepton in the context of lepton-specific two Higgs doublet models (2HDMs) with a local $U(1)'$ symmetry.  The SM particles are neutral under the $U(1)'$, but we assign appropriate $Z_2$ parities for both the SM and the vector-like lepton. Then, due to the $Z_2$ parity, there is a protection from Flavor Changing Neutral Currents with the leptophilic Higgs doublet.  The bare charged lepton mass terms are forbidden due to a nonzero $U(1)'$ charge of the leptophilic Higgs doublet in our model. Instead, the vector-like lepton carries a nonzero charge under the $U(1)'$ such that it mixes with the muon after the  dark Higgs and the leptophilic Higgs doublet get nonzero VEVs.  In this model, we show how the small muon mass is generated by a seesaw mechanism and the muon $g-2$ anomaly is explained  at the same time by the one-loop corrections with the vector-like lepton and $Z'$ dark gauge boson/dark Higgs boson.

The chirality violation with the vector-like lepton mass leads to an overall enhancement of the muon $g-2$ \cite{chirality,chirality-EFT,chirality3}, but it is compensated by the small mixing angles between the muon and the vector-like lepton. As a result, we show that the corrections to the muon $g-2$ are almost independent of the vector-like lepton mass, but there are non-decoupled effects from the vector-like lepton in the presence of light $Z'$/dark Higgs boson.
We also discuss the bounds from lepton flavor violating processes on the mixings of the other leptons with the vector-like lepton. For the consistency of the parameter region favored by the muon $g-2$, we also consider the constraints from electroweak precision data and Higgs data as well as collider searches.

The paper is organized as follows.
We begin with the model description for the vector-like lepton in lepton-specific 2HDMs with a local $U(1)'$ symmetry. Then, we discuss the seesaw mechanism for lepton masses and identify the $Z'$ and SM weak gauge interactions as well as the Yukawa interactions  for the vector-like lepton. Next, we show the new contributions to the muon $g-2$ in detail in our model and look for the parameter space for explaining the muon $g-2$ anomaly.
We also impose the bounds from lepton flavor violation on the other leptons and consider the constraints from electroweak precision and Higgs data, as well as the collider searches for $Z'$ and the vector-like lepton. 
Finally, conclusions are drawn.

\section{The model}

We introduce an $SU(2)_L$ singlet vector-like lepton, $E$, with charge $-2$ under the $U(1)'$  gauge symmetry \cite{exodm}.
We also have a dark Higgs field $\phi$ and a leptophilic Higgs doublet $H'$ with charges $-2$ and $+2$ under the $U(1)'$, respectively. We assume that the SM Higgs doublet $H$ and the SM fermions are neutral under the $U(1)'$.
Both $SU(2)_L$ singlet and doublet vector-like leptons have been introduced with or without a local $U(1)'$ symmetry in the literature \cite{VL1,VL2,VL3,VL4} in connection the muon $g-2$ anomaly, but bare lepton masses are present in those cases.

As in Type-X (or lepton-specific) two Higgs doublet models (2HDM) \cite{2HDM,2HDM2}, we also impose a $Z_2$ parity on the SM fields as well as new fields in our model.   Then, the lepton masses can be obtained from the mixing masses with the vector-like lepton, being proportional to the VEVs of the singlet scalar and the leptophilic Higgs doublet, but the flavor changing neutral currents at tree level is absent. The assignments for $U(1)'$ charges and $Z_2$ parities are given in Table~1.

\begin{table}[hbt!]
  \begin{center}
    \begin{tabular}{c|cccccccccc}
      \hline\hline
      &&&&&&&&&&\\[-2mm]
      & $q_L$ & $u_{R}$  &  $d_{R}$ & $l_{L}$  & $l_{R}$
      & $H$ & $H'$ & $E_L$ & $E_R$ & $\phi$  \\[2mm]
      \hline
      &&&&&&&&&&\\[-2mm]
      $U(1)'$ & $0$ & $0$ & $0$
            & $0$ & $0$ & $0$ & $+2$ & $-2$ & $-2$  & $-2$ \\[2mm]
          \hline
           &&&&&&&&&&\\[-2mm]
      $Z_2$ & $+$ & $-$ & $-$
            & $+$ & $+$ & $-$ & $+$ & $+$ & $+$ & $+$ \\[2mm]    
      \hline\hline
    \end{tabular}
  \end{center}
    \caption{$U(1)'$ charges and $Z_2$ parities.\label{charges}}
\end{table}

The Lagrangian for the SM Yukawa couplings including $Z'$, dark Higgs $\phi$, and the vector-like lepton is
\bea
{\cal L}=-\frac{1}{4} F'_{\mu\nu} F^{\prime \mu\nu} - \frac{1}{2} \sin\xi F'_{\mu\nu} B^{\mu\nu}+|D_\mu\phi|^2+|D_\mu H'|^2 - V(\phi,H,H') +{\cal L}_{VLSM}
\eea
with
\bea
{\cal L}_{VLSM}= -y_d {\bar q}_L d_R  H- y_u {\bar q}_L u_R {\tilde H} -M_E {\bar E}E-\lambda_E \phi {\bar E}_L  l_R-y_E  {\bar l}_L E_R H'  +{\rm h.c.}. \label{leptonL}
\eea
Here, ${\tilde H}=i\sigma^2 H^*$,  $F'_{\mu\nu}=\partial_\mu Z'_\nu-\partial_\nu Z'_\mu$, $B_{\mu\nu}$ is the field strength tensor for the SM hypercharge, the covariant derivatives are $D_\mu\phi=(\partial_\mu +2i g_{Z'} Z'_\mu)\phi$,  $D_\mu H'=(\partial_\mu -2i g_{Z'} Z'_\mu-\frac{1}{2}i g_Y  B_\mu- \frac{1}{2} i g \tau^i W^i_\mu) H'$.
Moreover,  $V(\phi,H,H')$ is the scalar potential respecting the $U(1)'$ symmetry for the singlet scalar $\phi$, the leptophilic Higgs $H'$ and the SM Higgs $H$, given by
\bea
V(\phi,H, H') &=& \mu^2_1 H^\dagger H + \mu^2_2 H'^\dagger H'  +(\mu_3 \phi H^\dagger H'+{\rm h.c.})  \nonumber \\
&&+ \lambda_1 (H^\dagger H)^2 + \lambda_2 (H'^\dagger H')^2+ \lambda_3 (H^\dagger H)(H'^\dagger H') \nonumber \\
&&+ \mu^2_\phi \phi^*\phi + \lambda_\phi (\phi^*\phi)^2+ \lambda_{H\phi}H^\dagger H\phi^*\phi +  \lambda_{H'\phi}H'^\dagger H'\phi^*\phi
\eea
where the $\mu_3$ term breaks the $Z_2$ parity softly.
From the fact that the second Higgs doublet $H'$ is charged under $U(1)'$ symmetry, we note that the quartic couplings for the Higgs doublets in 2HDMs  are constrained, and the mixing mass term between the two Higgs doublets is generated from the $\mu_3$ term after the $U(1)'$ symmetry is broken \cite{Bmeson}. 

We remark that the bare lepton masses are absent in eq.~(\ref{leptonL}) due to the nonzero $U(1)'$ charge for the leptophilic Higgs unlike the usual Type-X 2HDM. Instead, in our model, the physical lepton masses and the signal strength for the lepton Yukawa couplings to the SM Higgs can be generated correctly due to the mixing with the vector-like lepton, via the seesaw mechanism for leptons, as will be discussed in the next section.  As for the quark Yukawa couplings, the Yukawa couplings to the vector-like lepton can be flavor-dependent. But, the simultaneous presence of flavor-dependent couplings of the SM leptons to the vector-like lepton is subject to strong bounds for lepton flavor violating processes, such as $\mu\to e\gamma$. So, in general, we can introduce one vector-like lepton per generation for lepton masses without inducing the mixings between leptons.

After the dark Higgs gets a VEV as $\langle\phi\rangle=v_\phi$ and electroweak symmetry is broken by $\langle H\rangle=\frac{1}{\sqrt{2}}v_1$ and $\langle H'\rangle=\frac{1}{\sqrt{2}}v_2$, the squared mass of the $Z'$ gauge boson becomes $m^2_{Z'}=g^2_{Z'}(8 v^2_\phi+ 4 v^2_2)$. Moreover, we get the masses for electroweak gauge bosons as $m_Z=\frac{1}{2} \sqrt{g^2+g^2_Y}\,  v$ and $m_W=\frac{1}{2} g v$, with $v=\sqrt{v^2_1+v^2_2}$. The VEV of the SM Higgs doublet $H$ leads to quark masses and mixings, while the VEV of the extra Higgs doublet $H'$ leads to the mixing between the SM leptons and the vector-like lepton. 

Due to the gauge kinetic mixing and the nonzero $U(1)'$ charge of the leptophilic Higgs, there is a mass mixing between $Z$ and $Z'$ gauge bosons \cite{Bmeson}, which must be suppressed to satisfy the electroweak precision data and the collider bounds.
Moreover, there are similar electroweak precision bounds on the mixing angles between the leptons and the vector-like lepton.
In the following sections, we discuss the constraints from electroweak precision data and the collider searches in detail.

\section{Seesaw mechanism and new interactions for leptons}

We discuss the seesaw mechanism for generating small masses for charged leptons through the vector-like lepton and identify the $Z'$ and weak gauge interactions and the Yukawa interactions for the vector-like lepton and the SM leptons in our model.

\subsection{Seesaw mechanism for charged lepton masses}

Taking only the mixing between the vector-like lepton and one lepton, (which is muon for the later discussion on the muon $g-2$), we first enumerate the mass terms for the lepton sector as
\bea
{\cal L}_{L,{\rm mass}}&=& -M_E {\bar E}E -( m_R {\bar E}_L l_R+m_L {\bar l}_L E_R+ {\rm h.c.}) 
\eea
where $m_R, m_L$ are the mixing masses, given by $m_R=\lambda_E v_\phi$ and $m_L=\frac{1}{\sqrt{2}} y_E v_2$, respectively.
Then, after diagonalizing the mass matrix for leptons \cite{hmlee}, we can get the mass eigenvalues for leptons,
\bea
m^2_{l_1,l_2} = \frac{1}{2} \bigg( M^2_E +m^2_L +m^{ 2}_R\mp \sqrt{(M^2_E+m^{2}_L-m^2_R )^2+4m^2_RM^2_E}  \bigg),
\eea
and the rotation matrices for the right-handed leptons and the left-handed leptons are given by
\bea
\left(\begin{array}{c} l_L \\  E_L  \end{array}\right)&=&\left(\begin{array}{cc}  \cos\theta_L & \sin\theta_L\\   -\sin\theta_L & \cos\theta_L \end{array} \right)\left(\begin{array}{c} l_{1L}\\  l_{2L}  \end{array}\right), \\
\left(\begin{array}{c} l_R \\  E_R \end{array} \right)&=&\left(\begin{array}{cc}  \cos\theta_R & \sin\theta_R\\   -\sin\theta_R & \cos\theta_R \end{array} \right)\left(\begin{array}{c} l_{1R}\\  l_{2R} \end{array} \right),
\eea
with the mixing angles given by
\bea
\sin(2\theta_R) &=&\frac{2M_E m_R}{m^2_{l_2}-m^2_{l_1}}, \\
\sin(2\theta_L) &=&\frac{m^2_L}{m_{l_1} m_{l_2}}\,\cdot\sin(2\theta_R).  
\eea

It is remarkable that the mass squared for the physical charged lepton can be identified by $m^2_{l_1}$, which is approximated for $m_R, m_L\ll M_E$, as follows,
\bea
m^2_{l_1}\approx \frac{m^2_R m^2_L}{M^2_E}, \label{leptonmass}
\eea
whereas the mass squared for the vector-like lepton becomes
\bea
m^2_{l_2}\approx M^2_E+ m^2_L+m^2_R.
\eea
Therefore, a seesaw mechanism is at work for generating small masses for charged leptons due to heavy vector-like leptons.
For this purpose, from eq.~(\ref{leptonmass}), we need both the electroweak symmetry breaking with the leptophilic Higgs doublet for $m_L\neq 0$ and the $U(1)'$ symmetry breaking for $m_R\neq 0$ at the same time.
We note that the muon mass can be also generated radiatively by loop corrections, due to supersymmetric particles with soft terms \cite{muon-loop1} and a combination of vector-like leptons and extra scalars \cite{muon-loop2}.
But, in our case, the radiative corrections to the muon are protected by the $U(1)'$ symmetry.

Our result is reminiscent of the seesaw mechanism for neutrino masses, if we identify $E$ with heavy singlet neutrinos and $l$ with light active neutrinos. But, for the case of neutrinos,  the heavy singlet neutrinos do not have to be vector-like, so the resultant neutrino masses are of Majorana type, instead of Dirac type, unlike the case for charged leptons.  Experimental signatures of heavy neutrinos and vector-like leptons are also interesting in connection to the seesaw mechanism for neutrino masses \cite{HN}.

From eq.~(\ref{leptonmass}), we can explicitly write down the charged lepton mass in terms of the Yukawa couplings, as follows,
\bea
m_{l_1}\approx \frac{ \lambda_E y_E v_\phi v_2}{\sqrt{2} M_E}.
\eea
Then, choosing $m_{l_1}=m_\mu$ for the muon mass, the perturbativity conditions on the Yukawa couplings, $\lambda_E<1$ and $y_E<1$, set the upper limit on the vector-like mass by
\bea
M_E\simeq \frac{\lambda_E y_E v_\phi v_2}{\sqrt{2}  m_\mu}< 6700\,{\rm GeV} \bigg(\frac{v_\phi v_2}{10^3\,{\rm GeV}^2} \bigg).
\eea
Thus, the vector-like lepton can be decoupled from the weak scale, while generating the muon mass and satisfying the perturbativity conditions.

Moreover, for $m_R, m_L\ll M_E$, we also get the predictive results for the lepton mixing angles as
\bea
\sin(2\theta_R)&\simeq& \frac{2 m_{l_1}}{m_L},  \label{mixR} \\
\sin(2\theta_L) &\simeq & \frac{2m_L}{m_{l_2}}.  \label{mixL}
\eea
For instance, for $m_L\simeq \sqrt{m_{l_1} m_{l_2}}\simeq m_R$, the lepton mixings become comparable, $\theta_L\simeq \theta_R\simeq \sqrt{m_{l_1}/m_{l_2}}$. 
On the other hand, for $m_L\ll \sqrt{m_{l_1} m_{l_2}}$, we have a hierarchy of mixing angles, $\theta_L/\theta_R\simeq m^2_L/(m_{l_1} m_{l_2})\ll 1$, with $\theta_R\simeq m_{l_1}/m_L$.   We note that the product of the mixing angles is constrained to $\theta_L\theta_R\simeq m_{l_1}/m_{l_2}$, which depends only on the physical masses for lepton and vector-like lepton. In general, we can parametrize the small mixing mass parameters by $m_L\simeq(\theta_L/\theta_R)^{1/2} \sqrt{m_{l_1} m_{l_2}}$ and  $m_R\simeq(\theta_R/\theta_L)^{1/2} \sqrt{m_{l_1} m_{l_2}}$. 

We also note that the mixing angles between the leptons and the vector-like lepton can be hierarchical, due to the hierarchy of lepton masses. For instance, for electron and muon, the ratio of mixing angles becomes $(\theta^e_L \theta^e_R)/( \theta^\mu_L \theta^\mu_R)\simeq \frac{m_e}{m_\mu}\simeq 0.0048$ for the same $M_E$. Similarly, in the presence of the mixing between the tau lepton and the vector-like lepton, we would get the ratio of mixing angles as $(\theta^\mu_L \theta^\mu_R)/( \theta^\tau_L \theta^\tau_R)\simeq \frac{m_\mu}{m_\tau}\simeq 0.059$. But, as will be shown in the later section, the mixing angles for the electron and the tau lepton are strongly constrained by lepton flavor violating processes such as $\mu \to e \gamma$, so the mixing angles between the electron (or the tau lepton) and the vector-like lepton must be further suppressed. Thus, the masses for the electron and the tau lepton should be originated from different sources than the seesaw mechanism or the generalized seesaw mechanism with another vector-like lepton having a larger mass.

\subsection{Gauge interactions}

Consequently, including the gauge kinetic mixing and the mixing between the lepton and the vector-like lepton, we get  the approximate results for the effective interactions of leptons to $Z'$ and weak gauge bosons, as follows,
\bea
{\cal L}_{L,{\rm eff}} &= &-2 g_{Z'} Z'_\mu \Big(c^2_R {\bar E} \gamma^\mu P_R  E+ s^2_R\,  {\bar l} \gamma^\mu P_R l -s_Rc_R ({\bar E}\gamma^\mu P_R l + {\bar l}\gamma^\mu P_R E) \nonumber \\
&&\quad+c^2_L {\bar E} \gamma^\mu  P_L E+s^2_L {\bar l} \gamma^\mu P_L l -s_Lc_L ( {\bar E}\gamma^\mu P_L l + {\bar l}\gamma^\mu P_L E)   
  \Big) \nonumber \\
&&+\frac{g}{2c_W}\, Z_\mu (v_l+a_l) \Big( (c^2_L-1){\bar l} \gamma^\mu P_L l + s_Lc_L( {\bar E}\gamma^\mu P_L l+ {\bar l}\gamma^\mu P_L E)+s^2_L{\bar E} \gamma^\mu P_L E    \Big) \nonumber \\
&&+\frac{g}{2c_W}\, Z_\mu (v_l-a_l) \Big(  {\bar E} \gamma^\mu P_R  E+c^2_L {\bar E} \gamma^\mu P_L E + s^2_L{\bar l} \gamma^\mu P_L l - s_Lc_L( {\bar E}\gamma^\mu P_L l+ {\bar l}\gamma^\mu P_L E)   \Big) \nonumber \\
&&+\frac{g}{\sqrt{2}}\, W^-_\mu\Big( c_L{\bar l}\gamma^\mu P_L \nu+ s_L {\bar E} \gamma^\mu P_L \nu \Big) +{\rm h.c.} +{\cal  L}_{L,\xi},
\eea
with $s_R=\sin\theta_R, c_R=\cos\theta_R$, $s_L=\sin\theta_L, c_L=\cos\theta_L$, and  $v_l=\frac{1}{2}(-1+4s^2_W)$ and $a_l=-\frac{1}{2}$.
Here,  ${\cal  L}_{L,\xi}$ contains the extra couplings due to the gauge kinetic mixing,  given \cite{Bmeson} by
\bea
{\cal L}_{L,\xi}&=& Z'_\mu \bigg[e\, \xi c_\zeta c_W {\bar l} \gamma^\mu l +\frac{e}{2c_W s_W}(s_\zeta-t_\xi c_\zeta s_W) \Big( {\bar l} \gamma^\mu(v_l-a_l \gamma^5) l+{\bar \nu}\gamma^\mu P_L \nu \Big) \bigg] \nonumber \\
&&+  \frac{e}{2c_W s_W} (c_\zeta-t_\xi s_\zeta s_W) Z_\mu \Big[ {\bar l} \gamma^\mu(v_l-a_l \gamma^5) l+{\bar \nu}\gamma^\mu P_L \nu \Big],
\eea
where $c_\zeta=\cos\zeta, s_\zeta=\sin\zeta$, $t_\xi=\tan\xi$, and the mixing angle $\zeta$ between $Z$ and $Z'$ gauge bosons \cite{Bmeson} is given by
\bea
\tan 2\zeta =\frac{2m^2_{12} (m^2_{Z_2}-m^2_Z)}{(m^2_{Z_2}-m^2_Z)^2-m^4_{12}},
\eea
with  $m_Z$ being the $Z$-boson mass in the SM, $m_{Z_2}$ being the mass eigenvalue for the $Z'$-like gauge boson, and $m^2_{12}$ being the mixing mass given by
\bea
m^2_{12}=\frac{m^2_Z s_W}{c_\xi} \bigg( s_\xi -  \frac{4g_{Z'} v^2_2}{g_Y v^2}\bigg).
\eea
Here, we used the same notations for mass eigenstates as for interaction eigenstates, as they are lepton-like and vector-like.

We remark that the relevant effective vector and axial-vector couplings to $Z'$ and $Z$ containing both the leptons and the vector-like lepton  are given by
\bea
{\cal L}_{L,{\rm eff}}&\supset& g_{Z'} Z'_\mu {\bar l}\gamma^\mu (v'_l+a'_l \gamma^5) l + \Big(g_{Z'}Z'_\mu {\bar l} \gamma^\mu (c_V+c_A \gamma^5)E +{\rm h.c.}\Big) \nonumber \\
&&+\frac{g}{2c_W}\,(a_l s_L c_L)\, {\bar l} \gamma^\mu(1-\gamma^5)E Z_\mu +{\rm h.c.}
\eea
where
\bea
v'_l&=& e \varepsilon/g_{Z'}- \sin^2\theta_R-\sin^2\theta_L,  \label{ve} \\
a'_l &=& \sin^2\theta_L-\sin^2\theta_R, \label{ae} \\
c_V&=& \frac{1}{2} (\sin2\theta_R+\sin2\theta_L), \label{cv}  \\
c_A &=& \frac{1}{2} (\sin2\theta_R-\sin2\theta_L), \label{ca}
\eea
with $\varepsilon\equiv \xi\cos\theta_W$.
The $Z'$ interactions for transitions between the lepton and the vector-like lepton, explaining the muon $g-2$ anomaly by one-loop corrections with such unsuppressed transitions.  The mixing angle between the electron and the vector-like lepton is further suppressed by the electron mass as in eqs.~(\ref{mixR}) and (\ref{mixL}), so we can allow for the flavor-dependent couplings between leptons and $Z'$.

\subsection{Yukawa interactions}

We now consider the Yukawa interactions for neutral scalars and the charged Higgs in our model.

First, we expand the scalar fields around the VEVs, as follows,
\bea
H=\left(\begin{array}{cc} \phi^+_1 \\ \frac{1}{\sqrt{2}}(v_1+\rho_1+i\eta_1) \end{array} \right), \quad H'=\left(\begin{array}{cc} \phi^+_2 \\ \frac{1}{\sqrt{2}}(v_2+\rho_2+i\eta_2) \end{array} \right)
\eea
and
\bea
\phi=v_\phi+\frac{1}{\sqrt{2}}(\varphi+i\,a).
\eea
Then, we can identify two would-be neutral  Goldstone bosons, $G, G'$, and the CP-odd scalar $A$ \cite{Bmeson} as
\bea
G &=& \cos\beta\, \eta_1 +\sin\beta\,\eta_2,  \label{cpodd1} \\
G' &=& \frac{1}{\sqrt{2v^2_\phi +v^2_2}} \Big(\sqrt{2} v_\phi\, a-v\sin\beta\,\eta_2\Big),  \label{cpodd2}  \\
A&=& N_A \bigg( \sin\beta\, \eta_1-\cos\beta\,\eta_2 -\frac{v}{\sqrt{2}v_\phi}\,\sin\beta\cos\beta\, a \bigg)   \label{cpodd3} 
\eea
with
\bea
N_A= \frac{1}{\sqrt{1+v^2\sin^2\beta\cos^2\beta/(2v^2_\phi)}}.
\eea
Here, we note that $v_1=v\cos\beta$ and $v_2=v\sin\beta$.
Moreover, ignoring the mixing between the dark Higgs $\varphi$ and $\rho_{1,2}$, we also obtain the mass eigenstates for CP-even scalars, $h$ and $H$, as
\bea
h &=& \cos\alpha\,\rho_1 +\sin\alpha\, \rho_2, \\
H &=& -\sin\alpha\,\rho_1 + \cos\alpha\, \rho_2
\eea
where $\alpha$ is the mixing angle between CP-even scalars.

We also get the would-be charged Goldstone boson  $G^+$and the charged Higgs $H^+$ by
\bea
G^+ &=& \cos\beta\, \phi^+_1 +\sin\beta\,\phi^+_2, \\
H^+ &=& \sin\beta\,\phi^+_1 -\cos\beta\, \phi^+_2.
\eea

Therefore, we get the Yukawa couplings for quarks and leptons with the SM Higgs, and extra neutral scalar fields and the charged scalar,  as follows \footnote{We used the notations for extra Higgs scalars and their mixing angles in Ref.~\cite{Bmeson}},
\bea
{\cal L}_Y={\cal L}_q+{\cal L}_l
\eea
where
\bea
{\cal L}_q&=& - \sum_{f=u,d} \frac{m_f}{v} \,\bigg(\xi^f_h {\bar f} f  h+\xi^f_H {\bar f} f H -i\xi^f_A {\bar f} \gamma_5 f A \bigg) \nonumber \\
&&+\frac{\sqrt{2}}{N_A\, v}\, V_{ud}\, {\bar u} \Big(m_u\, \xi^u_A P_L +m_d\, \xi^d_A P_R \Big) d\, H^++{\rm h.c.}, \label{effYq}
\eea
with
\bea
\xi^u_h &=& \xi^d_h = \frac{\cos\alpha}{\cos\beta}, \\
\xi^u_H &=& \xi^d_H = -\frac{\sin\alpha}{\cos\beta}, \\
\xi^u_A &=& -\xi^d_A =N_A \tan\beta,
\eea
and
\bea
{\cal L}_l &=&-\frac{m_L}{v}\, (c_L{\bar l}+s_L {\bar E}) \Big( \xi^l_h P_R h + \xi^l_H P_R H-i \xi^{l_L}_A P_R A\Big)(c_R E- s_R l) +{\rm h.c.} \nonumber \\
&& +\frac{m_L}{v}\,  i \,\xi^{l_R}_A\, (c_R{\bar l}+s_R{\bar E}) P_L (c_L E-s_L l)  A+{\rm h.c.} \nonumber \\
&&-\frac{m_R}{\sqrt{2} v_\phi}\, (c_R {\bar l}+s_R {\bar E}) P_L (c_LE-s_L l)\varphi +{\rm h.c.} \nonumber \\
&& +\frac{\sqrt{2}m_L}{N_A \,v} \, {\bar \nu}\, \xi^{l_L}_A P_R (c_R E-s_R l) H^+ +{\rm h.c.}
\label{effYukawa}
\eea
with
\bea
\xi^l_h &=& \frac{\sin\alpha}{\sin\beta}, \\
\xi^l_H &=&  \frac{\cos\alpha}{\sin\beta}, \\
\xi^{l_L}_A &=& N_A \cot\beta, \\
\xi^{l_R}_A &=& -\frac{1}{2}N_A \Big(\frac{m_R}{m_L}\Big) \Big( \frac{v}{v_\phi}\Big)^2 \sin\beta\cos\beta.
\eea
Here, we used the same notations for mass eigenstates for leptons as for interaction eigenstates.
We note that $m_R, m_L, \theta_L, \theta_R$  are not independent parameters, but they are related from eqs.~(\ref{mixR}), (\ref{mixL}) and (\ref{leptonmass}), approximately given by
\bea
m_R\approx \frac{m_l}{\theta_L}, \quad m_L\approx \frac{m_l}{\theta_R}, \quad \theta_R \theta_L \approx \frac{m_l}{M_E},
\eea
with $m_l$ being the SM lepton mass.

In the alignment limit with $\alpha=\beta$, we can get the Yukawa couplings for the quarks as in the SM, namely, $\xi^u_h=\xi^d_h=1$.
 On the other hand, in order to get the correct lepton Yukawa couplings for the SM Higgs $h$ in the alignment limit with $\xi^l_h=1$,  from eq.~(\ref{effYukawa}), we need to take
$\cos\theta_L \sin\theta_R =\frac{m_l}{m_L}$ for the same branching ratio for $h\to \mu{\bar\mu}$ as in the SM \cite{hmumu}.
 This is consistent with the approximate mixing angle  for the vector-like lepton obtained for the correct muon mass by seesaw mechanism, namely, $\theta_R\simeq \frac{m_l}{m_L}$, for $\theta_R,\theta_L\ll 1$, obtained in eq.~(\ref{mixR}). 
 But, we note that the Yukawa coupling for the lepton has a wrong sign as compared to the SM, so it would be interesting to test this scenario through interference effects, such as $\mu{\bar \mu}\to hh$ at muon colliders \cite{dermisek}.
 
 We remark that the effective Yukawa interactions  for neutral scalar fields,  $h_i=h, H, \varphi, A$, and charged scalar $H^-$, containing the leptons and/or the vector-like leptons, which are relevant for the muon $g-2$, are given by
 \bea
 {\cal L}_l &\supset&  - \Big({\bar \mu}  (v^E_i- ia^E_i \gamma^5 )E  h_i +{\rm h.c.}\Big) -{\bar \mu}  (v^\mu_i- ia^\mu_i \gamma^5 ) \mu  h_i  \nonumber \\
 &&- \Big(v_{H^-} {\bar \mu}  (1-  \gamma^5 )\nu_\mu H^- +{\rm h.c.}\Big)
\eea
where
\bea
v^E_1&=& \frac{m_L}{2v}\,(c_L c_R-s_L s_R)\xi^\mu_h, \\
-ia^E_1&=&  \frac{m_L}{2v}\,(c_L c_R+s_L s_R)\xi^\mu_h, \\
v^E_2 &=&  \frac{m_L}{2v}\,(c_L c_R-s_L s_R)\xi^\mu_H, \\ 
 -ia^E_2&=& \frac{m_L}{2v}\,(c_L c_R+s_L s_R)\xi^\mu_H , \\
v^E_3 &=&  \frac{m_R}{2\sqrt{2} v_\phi}(c_L c_R-s_L s_R), \\
-i a^E_3&=& -\frac{m_R}{2\sqrt{2} v_\phi}(c_L c_R+s_L s_R), \\
v^E_4 &=& - \frac{i m_L}{2v}\,(c_L c_R+s_L s_R)(\xi^{\mu_L}_A+\xi^{\mu_R}_A),\\
a^E_4 &=&\frac{m_L}{2v}\,(c_L c_R-s_L s_R)(\xi^{\mu_L}_A-\xi^{\mu_R}_A), 
\eea
and
\bea
v^\mu_1 &=&-\frac{m_L}{v}\,c_L s_R\, \xi^\mu_h, \\
v^\mu_2 &=&-\frac{m_L}{v}\,c_L s_R\, \xi^\mu_H, \\
v^\mu_3 &=&-\frac{m_R}{\sqrt{2}v_\phi}\,c_R s_L,  \\ 
a^\mu_4 &=& -\frac{m_L}{v} (c_L s_R\, \xi^{\mu_L}_A -c_R s_L \, \xi^{\mu_R}_A), \\
v_{H^-} &=& \frac{\sqrt{2}m_L}{2 N_A\, v}\,s_R\,\xi^{\mu_L}_A,
\eea
and $v^\mu_4=a^\mu_{1,2,3}=0$

\section{Muon $g-2$ and lepton flavor violation} 

We present the details for the new contributions to the muon $g-2$ in our model and show the parameter space for explaining the muon $g-2$ anomaly.  We emphasize the dominant contributions coming from the lepton to vector-like lepton changing interactions with $Z'$ and singlet-like scalar $\varphi$, which remains nonzero in the decoupling limit of the vector-like lepton. 
We also discuss the constraints from flavor violating decays of leptons, in particular, $\mu\to e\gamma$, on the mixing angle between the electron and the vector-like lepton.

\subsection{Muon $g-2$}

From the combined average with Brookhaven E821 \cite{amu-exp,amu}, the difference from the SM value  becomes
\bea
\Delta a_\mu = a^{\rm exp}_\mu -a^{\rm SM}_\mu =251(59)\times 10^{-11}, \label{amu-recent}
\eea
which is now a $4.2\sigma$  discrepancy from the SM \cite{fermilab}.

Furthermore, the differences between the SM prediction for the anomalous magnetic moment of electron \cite{ae} and the experimental measurements in the cases of Cs atoms \cite{ae-exp} and Rb atoms\cite{Rb}, respectively, are given by
\bea
\Delta a_e &=& a^{\rm exp}_e -a^{\rm SM}_e =-88(36)\times 10^{-14}, \quad ({\rm Cs}) \label{ae-cs} \\
\Delta a_e &=& a^{\rm exp}_e -a^{\rm SM}_e =48(30)\times 10^{-14}, \quad ({\rm Rb}), \label{ae-rb}
\eea
showing $-2.4\sigma$ and $1.6\sigma$ deviations from the SM prediction.
The Rb data for fine structure constant shows an improvement in precision by factor 2, but it is in a tension with the Cs data at $5.4\sigma$, so there is a need of confirmation on those measurements.

\begin{figure}[t]
\centering
\includegraphics[width=0.27\textwidth,clip]{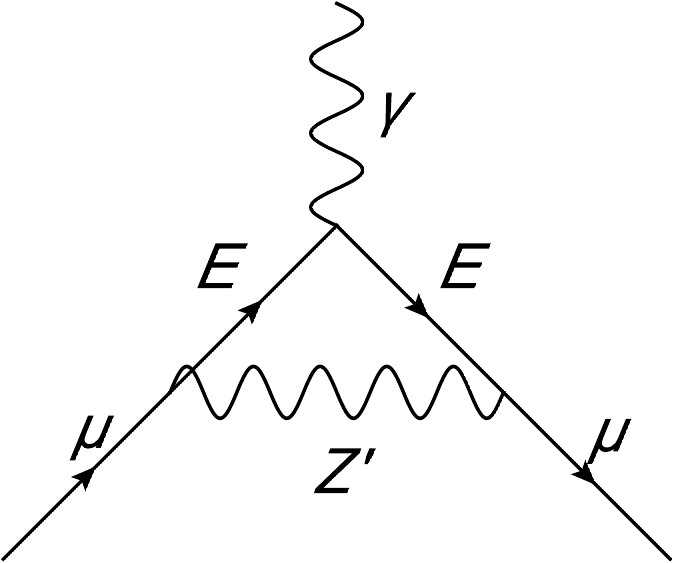}\,\,\,\,\,\,\,\,\,\,\,\,\,
\includegraphics[width=0.27\textwidth,clip]{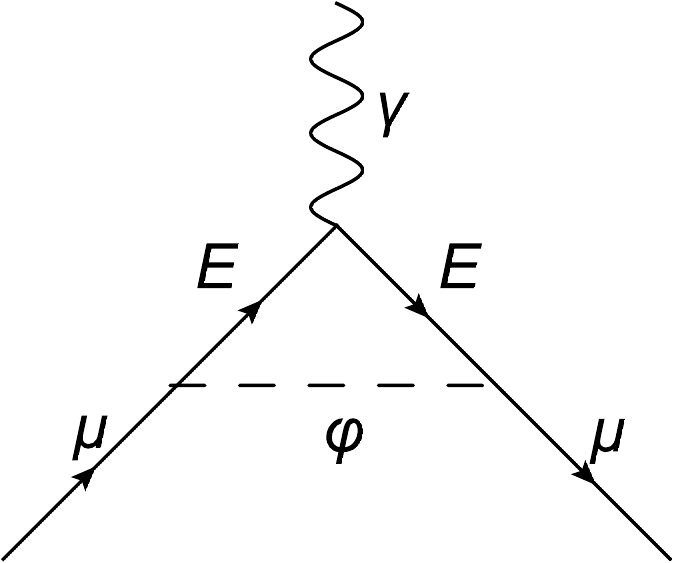}
\caption{
One-loop Feynman diagrams contributing to $\Delta a_{\mu}$, involving the vector-like lepton $E$
with $Z'$ on left and the dark Higgs $\varphi$ on right, respectively
}
\label{fig:g-2_diagram}
\end{figure}

In our model, the vector-like lepton, $Z'$ gauge boson as well as extra scalars contribute to the muon $g-2$ at one loop, as follows,
\bea
\Delta a_\mu= \Delta a^{Z',E}_\mu+\Delta a^{Z,E}_\mu+ \Delta a^{Z',\mu}_\mu  +\Delta a^{h,E}_\mu +\Delta a^{h,\mu}_\mu +\Delta a^{H^-}_\mu.
\eea
Unlike the Type-X (or lepton-specific) 2HDM or in models with vector-like leptons, the transitions between the SM lepton and the vector-like lepton via $Z'$ or $\varphi$, as shown in Fig.~\ref{fig:g-2_diagram}, give rise to the dominant contributions to the muon $g-2$ by $ \Delta a^{Z',E}_\mu$ and $\Delta a^{h,E}_\mu$. 
On the other hand, the Barr Zee-type contributions to $\Delta a_{\mu}$ at two-loops are suppressed by $ g^2_{Z'} \theta^2_{L,R}$ in addition to the extra loop factor, so we ignore the  2-loop contributions in the following analysis on the muon $g-2$.

We enumerate the one-loop corrections to the muon $g-2$ in our model in detail. 
First, we list $ \Delta a^{Z',E}_\mu $ as the contribution from $Z'$ and the vector-like lepton running together in loops  \cite{formula}, given by
\bea
\Delta a^{Z',E}_\mu &=&\frac{ g^2_{Z'} m^2_\mu}{4\pi^2} \int^1_0 dx \,  \bigg[ c^2_V \Big\{ x(1-x) \Big(x+\frac{2M_E}{m_\mu}-2 \Big)  \nonumber \\
&&-\frac{1}{2m^2_{Z'}} \Big( x^3(M_E-m_\mu)^2 + x^2 (M^2_E-m^2_\mu) \Big(1-\frac{M_E}{m_\mu} \Big) \Big)\bigg\}  +c^2_A \{ M_E\to -M_E\}\bigg] \nonumber \\
&&\times \Big(m^2_\mu x^2 + m^2_{Z'} (1-x) +x(M^2_E-m^2_\mu)\Big)^{-1} \label{amu0}
\eea
where $c_V, c_A$ are given in eqs.~(\ref{cv}) and (\ref{ca}).
The above contribution is given in the  closed form, as follows,
\bea
\Delta a^{Z',E}_\mu &=& \frac{ g^2_{Z'} m_\mu}{4\pi^2 m_{Z'}}r_E \bigg[ c^2_V \bigg\{ \frac{m_\mu}{m_{Z'}} \Big( \frac{5}{6} -\frac{5}{2}r_E+r^2_E +(r^3_E-3r^2_E+2r_E) \ln \frac{r_E-1}{r_E}\Big) \nonumber \\
&&  +\frac{M_E}{m_{Z'}}  \Big(2r_E-1 +2(r^2_E-r_E)  \ln \frac{r_E-1}{r_E}\Big)  \nonumber \\
&&+\frac{m_\mu M^2_E }{2m^3_{Z'}} \Big(\frac{5}{6} +\frac{3}{2} r_E +r^2_E +(r^2_E+r^3_E) \ln \frac{r_E-1}{r_E}  \Big) \nonumber \\
&& -\frac{M^3_E}{2m^3_{Z'}} \Big( \frac{1}{2} +r_E+r^2_E\ln \frac{r_E-1}{r_E} \Big) \bigg\} +c^2_A \{M_E\to -M_E \} \bigg] \label{amu}
\eea
with
\bea
r_E=\bigg(1-\frac{M^2_E}{m^2_{Z'}} \bigg)^{-1}.
\eea
Here, we approximated the masses for the lepton and the vector-like lepton to $m_{l_1}\simeq m_\mu$ and $m_{l_2}\simeq M_E$, respectively.
Then, we note that  the results can be approximated to
\bea
\Delta a^{Z',E}_\mu \simeq  \left\{\begin{array}{c} \frac{g^2_{Z'} M_E m_\mu}{16 \pi^2 m_{Z'}^2}\, (c_V^2-c_A^2), \qquad M_E\gg m_{Z'}, \vspace{0.3cm} \\ \frac{g^2_{Z'} M_E m_\mu}{4 \pi^2 m_{Z'}^2}\, (c_V^2-c_A^2), \quad m_\mu\ll M_E\ll m_{Z'}. \end{array}  \right. \label{zp1}
\eea
As a result, from eqs.~(\ref{cv}) and (\ref{ca}), we find that $c_V>c_A$, which leads to $\Delta a^{Z',E}_\mu>0$.
Since $c_V\simeq \theta_R+\theta_L$ and $c_A\simeq \theta_R-\theta_L$, with $\theta_R\simeq \frac{m_\mu}{m_L}$ and $\theta_L\simeq  \frac{m_L}{M_E}$, we find that $c_V^2-c_A^2\simeq 4\theta_L \theta_R\simeq \frac{4 m_\mu}{M_E}$. Therefore, we find that the large  chirality-flipping effect from $M_E$ is cancelled by the small mixing angles such that $\Delta a^{Z',E}_\mu\propto g^2_{Z'} m^2_\mu/m^2_{Z'}$ and it shows a non-decoupling phenomenon, almost independent of $M_E$.

Similarly, $\Delta a^{Z,E}_\mu$ is the contribution coming from the $Z$-boson and the vector-like lepton, which is given by eq.~(\ref{amu}), after the following replacements,
\bea
g_{Z'} &\to& \frac{g}{2c_W}, \quad m_{Z'} \to m_Z, \nonumber \\
c_V &\to&a_l s_L c_L=-\frac{1}{4} \sin 2\theta_L, \nonumber \\  
c_A &\to&\frac{1}{4} \sin 2\theta_L.
\eea
Then, for $M_E\gg m_Z$, we find the approximate results, as follows,
\bea
\Delta a^{Z,E}_\mu\simeq  -\frac{5g^2 c^2_V m^2_\mu}{ 96 \pi^2 c^2_W m_{Z}^2}\,\Big(1+\frac{6 m^2_{Z}}{5 M^2_E} \Big),
\eea
which is negative but subdominant for the small mixing angles for the vector-like lepton.

Likewise, $ \Delta a^{Z',\mu}_\mu $ is the contribution from $Z'$ and muon running together in loops, given by eq.~({\ref{amu0}), after the following replacements,
\bea
M_E\to m_\mu,\quad  c_V\to v'_\mu, \quad  c_A \to a'_\mu
\eea
where $v'_\mu, a'_\mu$ are given in eqs.~(\ref{ve}) and (\ref{ae}), respectively.
That is, we obtain 
\bea
\Delta a^{Z',\mu}_\mu&=& \frac{g^2_{Z'} m^2_\mu}{4\pi^2} \int^1_0 dx \,  \frac{\big[v^{\prime 2}_\mu x^2(1-x)-a^{\prime 2}_\mu \big(x(1-x)(4-x) +\frac{2m^2_\mu x^3}{m^2_{Z'}}\big)\big]}{m^2_\mu x^2+m^2_{Z'}(1-x)} \nonumber \\
&\simeq &\frac{g^2_{Z'} m^2_\mu}{12\pi^2 m^2_{Z'}}\, ( v^{\prime 2}_\mu -5 a^{\prime 2}_\mu)
\eea
where we assumed  $m_{Z'}\gg m_\mu$ in the second line. Thus, we find that $ \Delta a^{Z',\mu}_\mu $ is sub-dominant for a small gauge kinetic mixing and the small mixing angles for the vector-like lepton.

Moreover, $\Delta a^{h,E}_{\mu}$ are the contributions coming from the neutral scalars, $h_i=h, H, \varphi, A$, running together with the vector-like lepton in loops, given \cite{formula} by
\bea
\Delta a^{h,E}_\mu &=& \frac{m^2_\mu}{8\pi^2}  \sum_{i=1}^4\int^1_0 dx \,
\frac{\Big[|v^E_i|^2 \big(x^2-x^3+\frac{M_E}{m_\mu}\,x^2 \big)+|a^E_i|^2 \{M_E\to -M_E\} \Big]}{m^2_\mu x^2+(M^2_E-m^2_\mu)x+m^2_{h_i}(1-x)}. 
\eea
Then, we get the approximate results for the contributions from neutral scalars with vector-like lepton, as follows,
\bea
\Delta a^{h,E}_\mu \simeq \frac{m^2_\mu}{48\pi^2 M^2_E}   \Big[ |v^E_i|^2+|a^E_i|^2+ \frac{3M_E}{m_\mu}( |v^E_i|^2-|a^E_i|^2)  \Big], \,\, M_E\gg m_{h_i}, \label{scalar1}
\eea
or
\bea
\Delta a^{h,E}_\mu \simeq  \frac{m^2_\mu}{24\pi^2 m^2_{h_i}} \bigg[  |v^E_i|^2+|a^E_i|^2+\frac{3M_E}{m_\mu} ( |v^E_i|^2-|a^E_i|^2) \Big(\ln\Big(\frac{m^2_{h_i}}{M^2_E}\Big) -\frac{3}{2}\Big)\bigg],  \,\, M_E\ll m_{h_i}. \label{scalar2}
\eea
In either cases, the first part containing $|v^E_3|^2+|a^E_3|^2$ is comparable to the second part containing $ |v^E_3|^2-|a^E_3|^2$, because the latter chirality-enhanced contribution is compensated by the small mixing angles for the vector-like lepton. Namely, for the singlet-like scalar, $|v^E_3|^2+|a^E_3|^2\sim\frac{1}{4}(m_R/v_\phi)^2$  and $ |v^E_3|^2-|a^E_3|^2\sim -\frac{1}{2}(m_R/v_\phi)^2 s_L s_R\sim  -\frac{1}{2}(m_R/v_\phi)^2 \frac{m_\mu}{M_E}$.
 Similarly as for the loops with $Z'$ and the vector-like lepton, there is a  non-decoupling phenomenon, almost independent of $M_E$. 
In the decoupling limit for the neutral scalars in 2HDM, we find that the singlet-like scalar loops with vector-like lepton, $\Delta a^{\varphi,E}_\mu$, can contribute a sizable negative contribution to the muon $g-2$, as shown from eqs.~(\ref{scalar1}) and (\ref{scalar2}).

\begin{figure}[t]
\centering
\includegraphics[width=0.43\textwidth,clip]{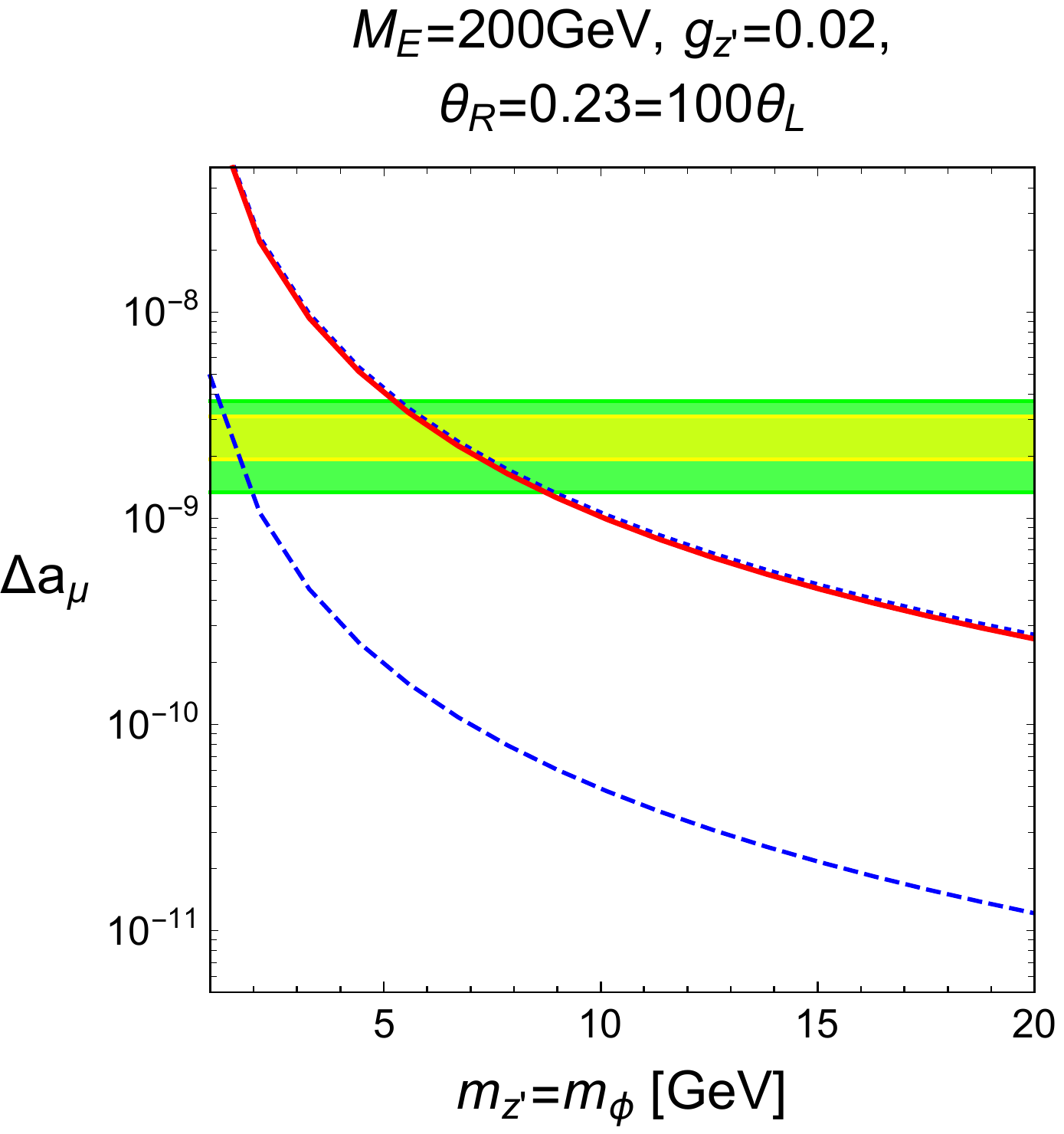} \,\,\,\,
\includegraphics[width=0.43\textwidth,clip]{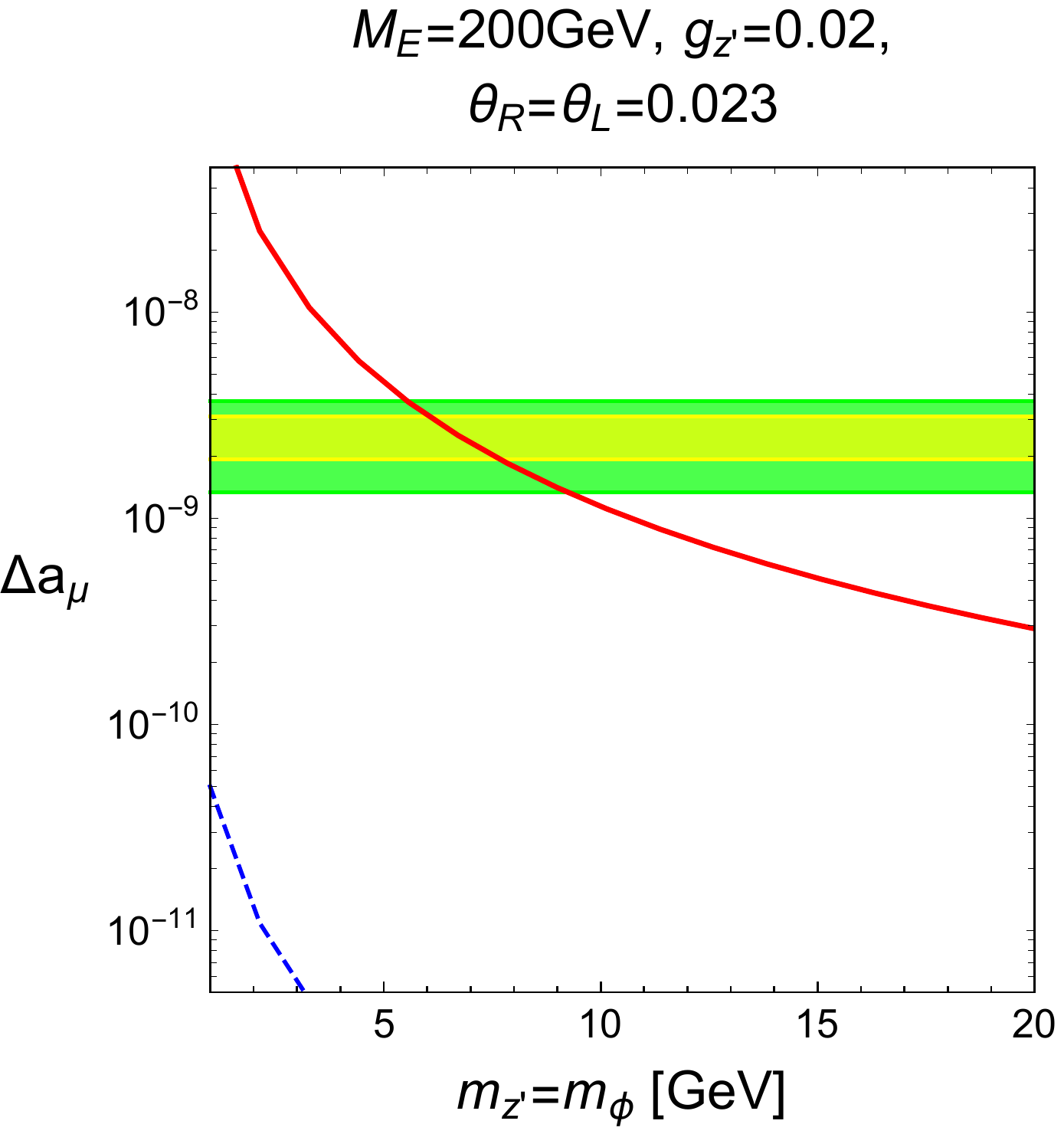} 
\caption{
$\Delta a_{\mu}$ from the 1-loop digrams in Fig.~\ref{fig:g-2_diagram} as a function of $m_{Z'}=m_{\varphi}$. 
The one-loop contribution from $Z'$ and vector-like lepton (from the left in Fig.~\ref{fig:g-2_diagram}), the one-loop contribution from the dark Higgs $\varphi$ and the vector-like lepton (from the right in Fig.~\ref{fig:g-2_diagram}), and the combined one-loop results, are shown in blue dotted, blue dashed, red solid line, respectively.
The yellow (green) bands indicate the deviations of the muon $g-2$ from the SM value within $1\sigma$($2\sigma$), after the $g-2$ results from both Fermilab  E989 and Brookhaven E821 are combined.
We took  $\theta_R=0.23=100\theta_L$ on left and $\theta_R=\theta_L=0.023$ on right. We chose  $M_E= 200 \,{\rm GeV}$, $g_{Z'}=0.02$, $m_{Z'}=m_{\varphi}$ and $v_\phi=m_{Z'}/(2\sqrt{2} g_{Z'})$ for both plots.
}
\label{fig:g-2}
\end{figure}

On the other hand, $\Delta a^{h,\mu}_\mu $ coming from the neutral scalars and muon in loops is given by
\bea
\Delta a^{h,\mu}_\mu &=& \frac{m^2_\mu}{8\pi^2}\, \sum_{i=1}^3 |v^\mu_i|^2 \int^1_0 dx \,
\frac{x^2(2-x)}{m^2_\mu x^2+m^2_{h_i}(1-x)} \nonumber \\
&& - \frac{m^2_\mu}{8\pi^2}\,|a^\mu_4|^2 \int^1_0 dx \,
\frac{x^3}{m^2_\mu x^2+m^2_{A}(1-x)}.
\eea
which is sub-dominant due to the suppression with the small mixing angles for the vector-like lepton.

Finally, $\Delta a^{H^-}_\mu$ is the contribution coming from the charged Higgs running in loops \cite{formula}, given by
\bea
\Delta a^{H^-}_\mu &=& \frac{m^2_\mu}{4\pi^2}\,|v_{H^-}|^2\int^1_0 dx \,\frac{x^3-x^2}{m^2_\mu x^2+(m^2_{H^-}-m^2_\mu) x+m^2_\mu(1-x) } \nonumber \\
&\simeq & -\frac{m^2_\mu}{24\pi^2 m^2_{H^-}}\, |v_{H^-}|^2,
\eea
which is doubly suppressed by the charged Higgs mass and the small mixing angles for the vector-like lepton.

In Fig.~\ref{fig:g-2}, from the one-loop graphs in Fig.~\ref{fig:g-2_diagram}, we depict $\Delta a_{\mu}$  as a function of $m_{Z'}=m_\varphi$.
We chose $M_E = 200$~GeV for both plots, and took $\theta_R= 0.23$, $\theta_L= 0.0023$, on left, which leads to $m_L= 0.46\,{\rm GeV}$ and $m_R= 46\,{\rm GeV}$,  and $\theta_R=\theta_L=0.023$, on right, which leads to $m_L=m_R= 4.6\,{\rm GeV}$.   The $Z'$ contributes positively to $\Delta a_{\mu}$ because the vectorial coupling is larger than the axial coupling in our model. On the other hand, the dark Higgs contributes negatively to $\Delta a_{\mu}$, because the pseudo-scalar coupling is larger than the scalar coupling. 
For $m_{Z'}=m_{\varphi}$ and  $g_{Z'}=0.02$, namely, $\lambda_\phi\simeq 2 g^2_{Z'}=0.0008$ for $v_2\ll v_\phi$, and $v_\phi=m_{Z'}/(2\sqrt{2} g_{Z'})=18 m_{Z'}$, we find that $Z'$ and dark Higgs masses around $5$--$8$~GeV are favored to explain the experimental value of the muon $g-2$. We find that for either $\theta_R\gg \theta_L$ or $\theta_R=\theta_L$, the one-loop corrections with $Z'$ and vector-like lepton contribute dominantly to the muon $g-2$.

\begin{figure}[t]
\centering
\includegraphics[width=0.43\textwidth,clip]{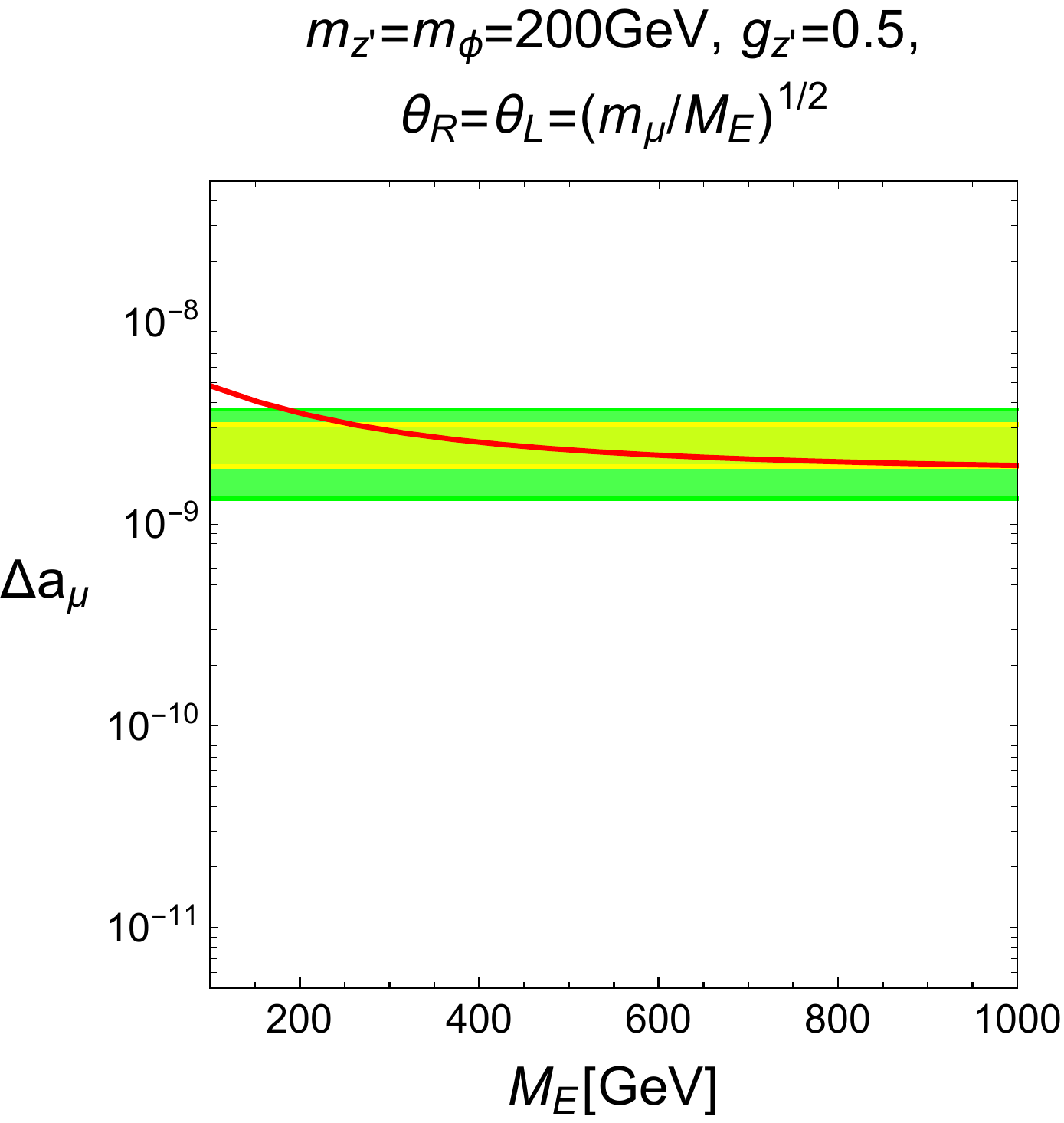}\,\,\,\,
\includegraphics[width=0.40\textwidth,clip]{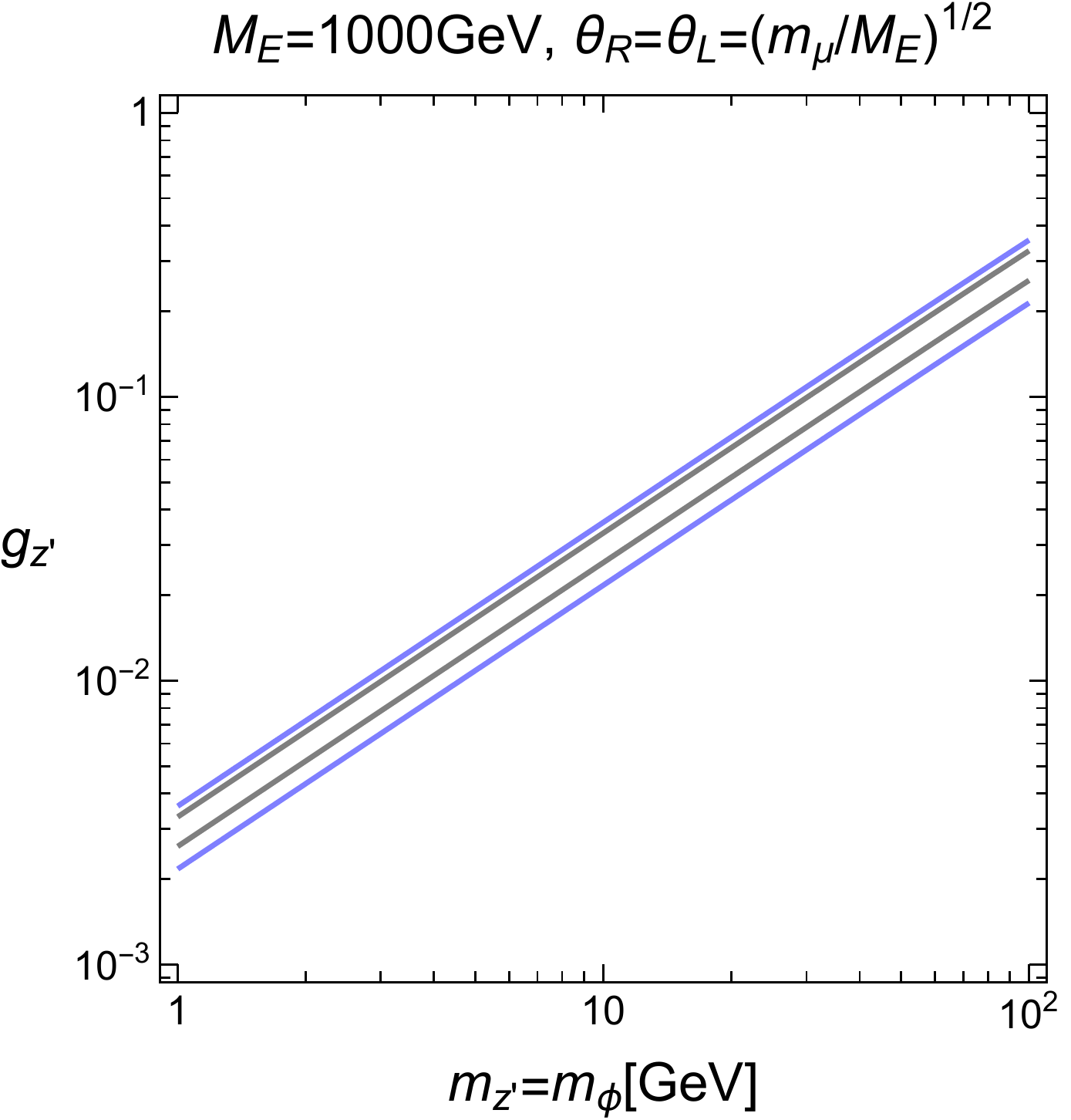}
\caption{(Left) $\Delta a_{\mu}$  as a function of $M_E$ in red line, in comparison to $1\sigma$($2\sigma$) bands for the deviation of the muon $g-2$ in yellow(green). We fixed $m_{Z'}=m_\phi=200\,{\rm GeV}$ and $g_{Z'}=0.5$, and  the vector-like mixing angles to $\theta_R=\theta_L=\sqrt{m_\mu/M_E}$.  (Right) The new contribution to the muon $g-2$ in the parameter space for $m_{Z'}=m_{\phi}$ versus $g_{Z'}$ within $1\sigma$($2\sigma$), shown between black(blue) lines.  We took $M_E=1000\,{\rm GeV}$ and $\theta_R=\theta_L=\sqrt{m_\mu/M_E}$.
}
\label{fig:g-2b}
\end{figure}

On the left of  Fig.~\ref{fig:g-2b}, we depict the new contribution to the muon $g-2$ as a function of the vector-like lepton mass $M_E$ in comparison to   $1\sigma$($2\sigma$) bands for the deviation from the SM value of the muon $g-2$ in yellow(green).  Here, we  chose  $m_{Z'}=m_{\varphi}=200\,{\rm GeV}$ and $g_{Z'}=0.5$ and  $\theta_R=\theta_L=\sqrt{m_\mu/M_E}$.  In this case, the mixing mass parameters are chosen to $m_R=m_L=\sqrt{m_\mu M_E}$, which varies between $3-10\,{\rm GeV}$ for $M_E=100-1000\,{\rm GeV}$. 
On the right of Fig.~\ref{fig:g-2b}, we also show the contours for the new contribution to the muon $g-2$ within $1\sigma$($2\sigma$) bands in the parameter space for  $m_{Z'}=m_{\phi}$ and $g_{Z'}$. In this case, we chose $M_E=1000\,{\rm GeV}$ and $\theta_R=\theta_L=0.01$.
For both plots, we find that the $Z'$ loops with vector-like lepton give rise to a dominant contribution to the muon $g-2$ and become independent of vector-like lepton masses for $M_E\gg m_{Z'}=m_{\phi}$.

\subsection{Bounds from lepton flavor violation}

In the presence of the simultaneous mixings of the vector-like lepton with muon and the other leptons (electron and tau), the one-loop diagrams with $Z'$ and the vector-like lepton contributes to the branching ratios of $\mu\rightarrow e \gamma$, $\tau\to \mu\gamma$ and $\tau\to e\gamma$. For $M_E\gg m_{Z'}$, we obtain the decay branching ratios as 
\bea
{\rm BR}(\mu\rightarrow e\gamma)&\simeq&\tau_\mu\,\cdot\frac{\alpha m_\mu^3 }{256\pi^4}\, \Big(\frac{g^4_{Z'} M^2_E}{m^4_{Z'}}\Big)\, [(\theta^\mu_R \theta^e_L)^2+(\theta^\mu_L \theta^e_R)^2], \\
{\rm BR}(\tau\rightarrow \mu \gamma)&\simeq&\tau_\tau\,\cdot\frac{\alpha m_\tau^3 }{256\pi^4}\, \Big(\frac{g^4_{Z'} M^2_E}{m^4_{Z'}}\Big)\, [(\theta^\tau_R \theta^\mu_L)^2+(\theta^\mu_L \theta^\tau_R)^2], \\
{\rm BR}(\tau\rightarrow e\gamma)&\simeq&\tau_\tau\,\cdot\frac{\alpha m_\tau^3 }{256\pi^4}\, \Big(\frac{g^4_{Z'} M^2_E}{m^4_{Z'}}\Big)\, [(\theta^\tau_R \theta^e_L)^2+(\theta^e_L \theta^\tau_R)^2]
\eea
where the lifetimes of muon and tau are given by $\tau_\mu=2.197\times 10^{-6}\, {\rm s}$ and $\tau_\tau=(290.3\pm 0.5)\times 10^{-15}\, {\rm s}$, respectively \cite{pdg}, and $\theta^l_{R,L}$ are the mixing angles of the lepton $l$ with the vector-like lepton.  
We also note that there are additional one-loop contributions to the above lepton flavor violating processes coming from to the extra scalar fields, $h_i$ with $i=h,H, \varphi$ and $A$, but they are negligible in the parameter space where the deviation for the muon $g-2$ is explained by the $Z'$ loop contributions. 

From the current experimental bound on lepton flavor violating decays, given \cite{meg,taumu} by 
\bea
{\rm BR}(\mu\rightarrow e \gamma)&<&4.2\times 10^{-13}, \label{mue} \\
{\rm BR}(\tau\rightarrow\mu\gamma)&<&4.4\times 10^{-8},  \label{taumu} \\
{\rm BR}(\tau\rightarrow e\gamma)&<&3.3\times 10^{-8}.
 \eea
we can constrain  the mixing angles between the other leptons and the  vector-like lepton.

We choose a set of the $Z'$ mass and coupling and the mixing between the muon and the vector-like lepton, favored by the muon $g-2$, as $m_{Z'}=6(200)\,{\rm GeV}$ and $g_{Z'}=0.02(0.5)$ and $\theta^\mu_L=\theta^\mu_R=\sqrt{m_\mu/M_E}$. Then, imposing the bound on $\mu\to e\gamma$ in eq.~(\ref{mue}), we find that the mixing angles for the electron are constrained to $\theta^e_R=\theta^e_L\lesssim 2 (3)\times 10^{-4}\sqrt{m_e/M_E}$, being suppressed more than the one inferred from the seesaw condition with one vector-like lepton, $\theta^e_R\simeq \sqrt{m_e/M_E}$.
Therefore, as the dominant contribution to the lepton $g-2$ scales by $\Delta a_l\sim g^2_{Z'} M_E m_l  (\theta^l_L \theta^l_R)/m^2_{Z'}$ from eq.~(\ref{zp1}),  the ratio of the electron $g-2$ to the muon $g-2$  contributions in our model is quite suppressed as $\Delta a_e/\Delta a_\mu\lesssim 4(9)\times 10^{-8} (m_e/m_\mu)^2$. 

Similarly, for the same set of parameters favored by the muon $g-2$ as in the previous paragraph, we find from the bound on $\tau\to \mu \gamma$  in eq.~(\ref{taumu}) that the mixing angles for the tau lepton are also constrained to $\theta^\tau_R=\theta^\tau_L\lesssim 0.04 (0.08)\sqrt{m_\tau/M_E}$, being suppressed again more than the one inferred from the seesaw condition with one vector-like lepton, $\theta^\tau_R\simeq \sqrt{m_\tau/M_E}$.
Therefore, in order to be consistent with lepton flavor violation, we find that the other lepton masses should be determined dominantly by their mixings with extra vector-like leptons.

\section{Precision and collider constraints}

In this section, we include the indirect constraints on the vector-like lepton from electroweak precision data and Higgs data and also consider the direct limits on the model from collider searches.

\subsection{Electroweak precision and Higgs data}

The global fit in PDG data \cite{pdg} shows $\Delta\rho=(3.9\pm 1.9)\times 10^{-4}$, which is $2\sigma$ above the SM expectation $\rho=1$, thus strongly constraining the gauge couplings of the vector-like lepton and the $Z'$ interactions.

The vector-like lepton and $Z'$ interactions also modify the $\rho$ parameter to
\bea
\Delta\rho &=&\Delta\rho_L+\Delta\rho_H
\eea
where $\Delta\rho_L$ is the contribution from the vector-like lepton, given \cite{lavoura,hmlee} by
\bea
\Delta\rho_L = \frac{\alpha}{16\pi s^2_W c^2_W}\, \sin^2\theta_L \bigg[\frac{M^2_E}{m^2_Z}-   \frac{m^2_\mu}{m^2_Z}-(\cos^2\theta_L) \theta_+(z_E,z_\mu) \bigg], \label{rhoL}
\eea
with $z_E =M^2_E/m^2_Z$, $z_\mu=m^2_\mu/m^2_Z$, and
\bea
\theta_+(a,b)= a+b -\frac{2ab}{a-b} \ln \frac{a}{b}, 
\eea
and $\Delta\rho_H$ is the contribution from $Z'$, given \cite{Bmeson} by
\bea
\Delta\rho_H=\frac{m^2_W}{m^2_{Z_1} \cos^2\theta_W}\,(\cos\zeta-\sin\theta_W\tan\xi\sin\zeta)^2-1.
\eea
Here,  $m_{Z_1}$ is the mass eigenvalue for the $Z$-like gauge boson. 

For the heavy vector-like lepton with $M_E\gtrsim m_Z$, we have $\theta_+(z_E,z_\mu)\simeq \frac{1}{m^2_Z}(M^2_E-4m^2_\mu\ln (M_E/m_\mu))$ in eq.~(\ref{rhoL}), so
we can approximate the vector-like lepton contributions, $\Delta\rho_L $, to
\bea
\Delta\rho_L \simeq  \frac{\alpha M^2_E}{16\pi s^2_W c^2_W m^2_Z} \,\sin^4\theta_L.
\eea
So, for $\Delta\rho_L \lesssim 10^{-4}$, we need $|\sin\theta_L|\lesssim 0.56\, (100\,{\rm GeV}/M_E)^{1/2}$. 
On the other hand, for light or heavy $Z'$, we quote the following approximate results for $\Delta\rho_H$,
\bea
\Delta\rho_H\simeq \left\{ \begin{array}{cc} \frac{s^2_W m^2_Z}{c^2_\xi m^2_{Z'} }\, \bigg(\frac{16g^2_{Z'}}{g^2_Y}\,\sin^4\beta-\sin^2\xi \bigg), \quad  m_{Z'}\gg m_{Z}, \vspace{0.3cm}\\ -\frac{s^2_W}{c^2_\xi}\, \Big(s_\xi-\frac{4 g_{Z'}}{g_Y}\,\sin^2\beta \Big) \Big(3s_\xi - \frac{4g_{Z'}}{g_Y}\,\sin^2\beta\Big), \quad m_{Z'}\ll m_Z. \end{array} \right.
\eea
Thus, in order to satisfy the current bound on $\Delta\rho$, it is sufficient to choose either $|\sin\xi|\lesssim 10^{-2}$ and $\sin\beta\lesssim 0.1 \sqrt{(g_Y/g_{Z'})}$ for $m_{Z'}\ll m_Z$ unless there is a cancellation. The bounds are less severe for $m_{Z'}\gg m_Z$, due to the overall suppression by $m^2_Z/m^2_{Z'}$.

Since the vector-like lepton couples to the SM Higgs boson in our model, it modifies the diphoton decay mode of the Higgs boson. Then, the ratio of the modified diphoton decay rate to the SM value is given \cite{hmlee,diphoton} by
\bea
R_{\gamma\gamma}= \left| 1+ \frac{\frac{m_L}{M_E}\,s_L c_R\, \xi^l_h A_f(\tau_E)}{A_V(\tau_W) + N_c Q^2_t A_f (\tau_t)}\right|^2
\eea
where $\tau_W=m^2_h/(4M^2_W)$, $\tau_t=m^2_h/(4m^2_t)$, $\tau_E=m^2_h/(4M^2_E)$, and the loop functions are given by
 \bea
 A_V(x) &=&-x^{-2} \Big[ 2x^{2} +3x +3(2x-1)f(x) \Big],  \\
 A_f(x)&=& 2x^{-2} \Big[x+(x-1)f(x) \Big]
 \eea
 with
 \bea
f(x) =\left\{\begin{array}{c}  {\rm arcsin}^2\sqrt{x}, \qquad\quad  x\leq 1, \vspace{0.2cm} \\  -\frac{1}{4}\bigg[\ln \frac{1+\sqrt{1-x^{-1}}}{1-\sqrt{1-x^{-1}}} -i\pi \bigg]^2, \quad x>1.  \end{array} \right.
\eea
In our model, the mixing angle for the left-handed leptons is approximated to $\theta_L\simeq m_L/M_E$ from eq.~(\ref{mixL}), so the extra contribution of the vector-like lepton to the Higgs diphoton rate is suppressed by $\theta^2_L$, namely, the diphoton rate changes by $R_{\gamma\gamma}-1\simeq -0.4\, \theta^2_L \,\xi^l_h$ for $M_E\gg m_h$. Therefore, in the alignment limit with $\xi^h_l=1$, it is sufficient to choose the mixing angle to $\theta_L\lesssim 0.5$ for $|R_{\gamma\gamma}-1|\lesssim 0.1$ from Higgs data.

\subsection{$Z'$ and vector-like lepton at colliders}

The vector-like lepton can be produced by the Drell-Yann processes with off-shell $\gamma^*$ and $Z^*$ in the $s$-channels at LEP and LHC. Since the vector-like lepton is an $SU(2)_L$ singlet in our model, the Drell-Yann production cross section with $Z^*$ is suppressed by $\sin^4\theta_W$.

For $M_E>m_Z, m_{Z'}$, the vector-like lepton can decay by $E\to W\,\nu, Z\, l, Z' l, \varphi\,l$ while  $Z'$ decays dominantly into a pair of muon and anti-muon.
For unitarity, the dark Higgs $\varphi$ has a comparable mass as the $Z'$ mass, so $\varphi$ decays into  a pair of muon and anti-muon. The dark Higgs $\varphi$ can decay into a photon pair through loops too.
For $M_E> m_{h_i}, m_{H^-}$, with $h_i=h, H, \varphi, A$, the vector-like lepton can decay by $E\to h_i \, l$ or $E\to H^-\,\nu$, but the additional decay modes of the vector-like lepton are suppressed by the mixing angles of the vector-like lepton. 

We also remark on the production and decays of $Z'$ in our model.
In the absence of the gauge kinetic mixing, $Z'$ couples only to the SM leptons through the mixing angles for the vector-like lepton, so the production cross section for $Z'$ at LEP is suppressed by the small mixing angles for the electron.
For a nonzero gauge kinetic mixing, $Z'$ also couples to the other leptons and the quarks in the SM so there are stringent bounds from the dimuon searches for $Z'$ at  LEP and LHC. In particular, the LHC dimuon searchers \cite{dimuon,kawamura} set the limit on the production cross section to $\sigma(pp\to Z'\to \mu{\bar\mu})\lesssim 10\,{\rm fb}$ for $m_{Z'}\gtrsim 250\,{\rm GeV}$ \cite{kawamura}. But, we can tolerate such dimuon bounds for $Z'$ with a small gauge kinetic mixing or a small $Z'$ mass.

If $M_E<m_{Z'}< 2M_E$, the $Z'$ gauge boson can decay by $Z'\to E\,{\bar l}, {\bar E}\, l$, becoming dominant decay channels.  
In this case, for $E\to Z\,l$, there can be at least two leptons in the final state from the $Z'$ decay.
For $m_{Z'}>2M_E$, the $Z'$ gauge boson can decay dominantly by $Z'\to E{\bar E}$, leading to at least four leptons in the final state.

\begin{figure}[t]
\centering
\includegraphics[width=0.45\textwidth,clip]{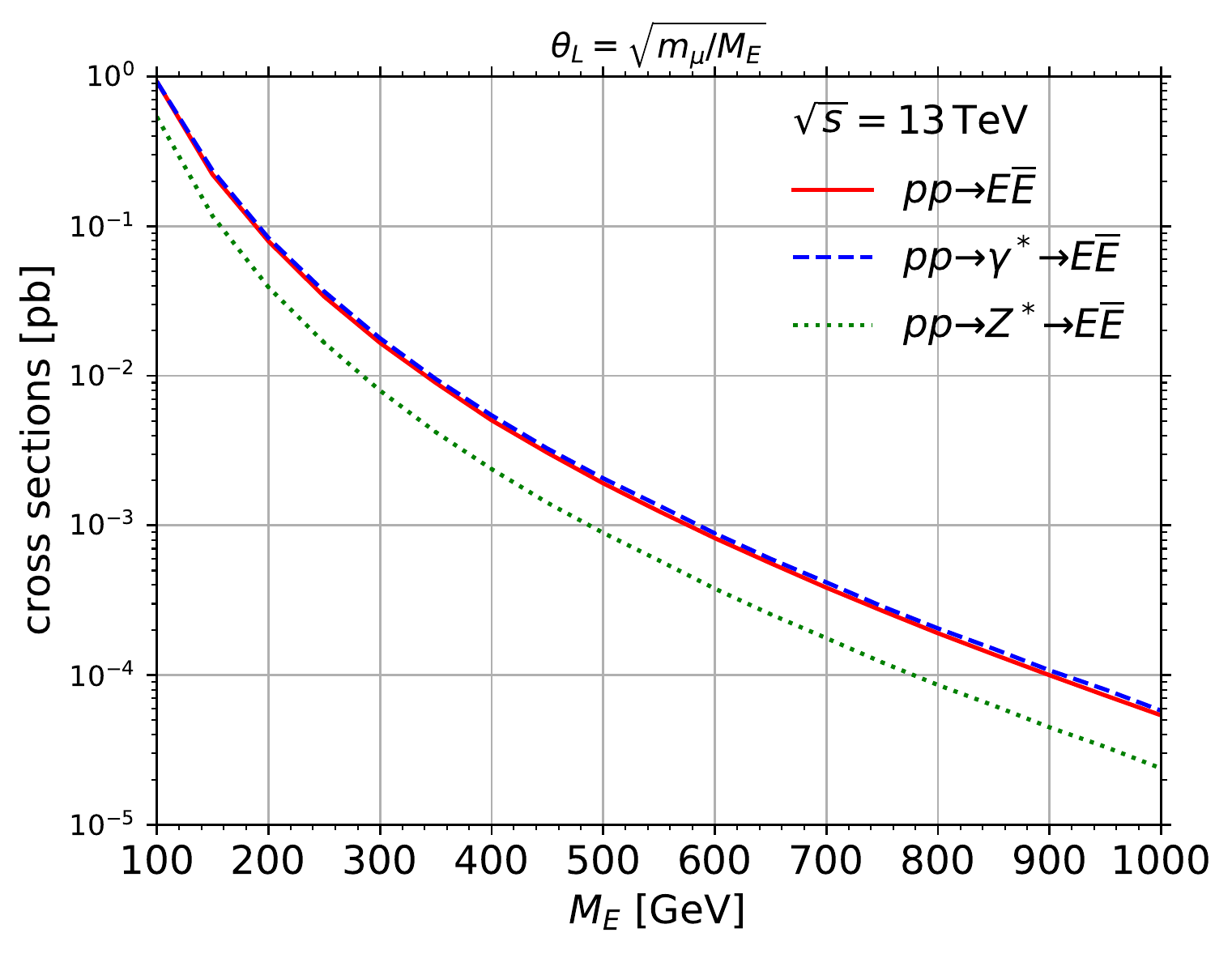} \,\,\,\,\,\,
\includegraphics[width=0.40\textwidth,clip]{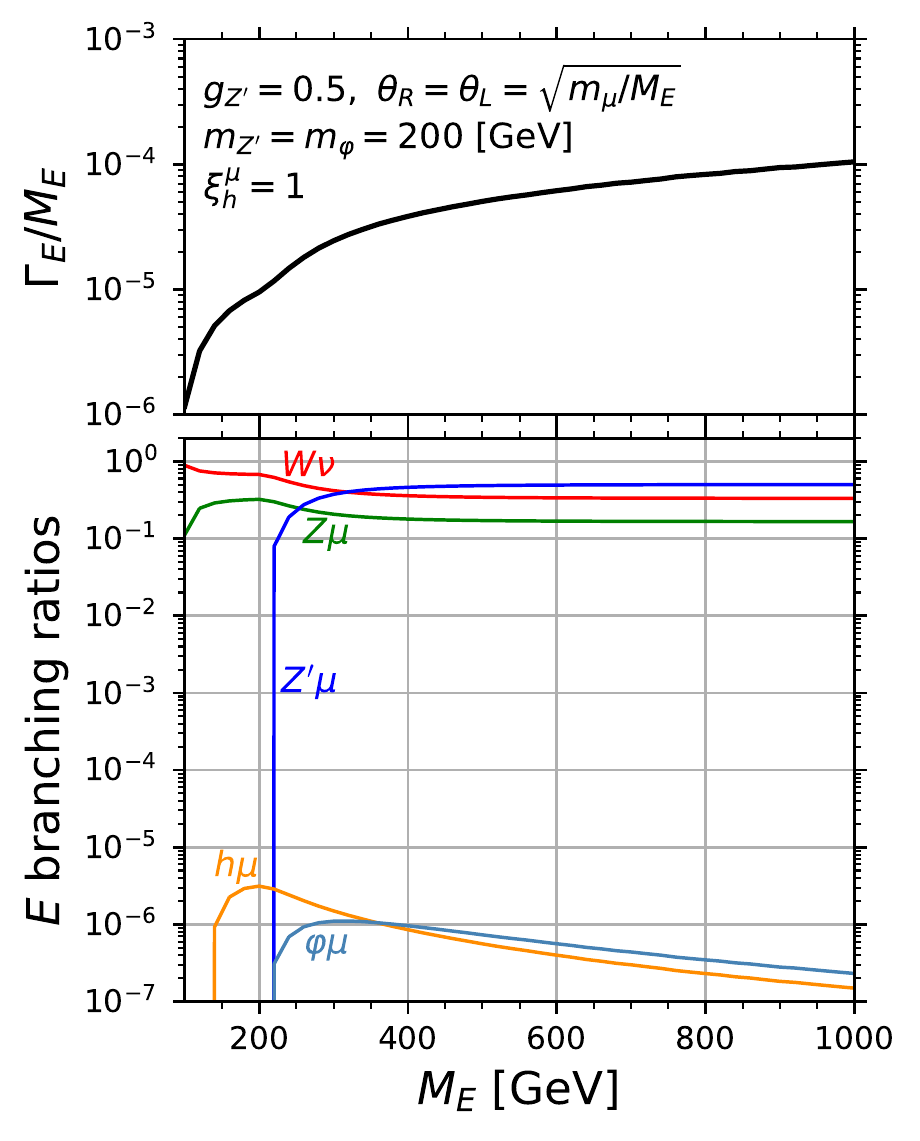} 
\caption{Production cross section and decay Branching Ratios for a vector-like lepton pair as a function of the vector-like lepton mass at the LHC $13\,{\rm TeV}$. We fixed the mixing angles for the vector-like lepton to $\theta_L=\sqrt{m_\mu/M_E}$. The total cross section and those with off-shell photon and off-shell $Z$-boson are show in red solid, blue dashed, green dotted lines, respectively.  For the decay BRs on right,  we also chose $g_{Z'}=0.5$, $m_{Z'}=m_\phi=200\,{\rm GeV}$, $\theta_R=\theta_L=\sqrt{m_\mu/M_E}$, and the alignment limit for the Higgs couplings, $\xi^\mu_h=1$.
}
\label{fig:VL-LHC}
\end{figure}

In the left plot of Fig.~\ref{fig:VL-LHC}, the production cross section for a pairs of vector-like leptons at the LHC $13\,{\rm TeV}$ is shown as a function of $M_E$. We have drawn the total production cross section and the partial production cross sections with  $\gamma^*$ and $Z^*$ in red solid, blue dashed, green dotted lines, respectively.  We have set $\theta_L=\sqrt{m_\mu/M_E}$. 
 We also depicted the decay Branching Ratios(BRs) of the vector-like lepton with its total decay rate in the upper panel, in the right plot of Fig.~\ref{fig:VL-LHC}. Here, we have also chosen  $g_{Z'}=0.5$, $m_{Z'}=m_\phi=200\,{\rm GeV}$, and $\theta_R=\theta_L=\sqrt{m_\mu/M_E}$, and assumed the alignment limit for the Higgs Yukawa couplings  by $\xi^\mu_h=1$. We also took $m_H, m_A, m_{H^-}>M_E$. 

For light $Z'$, the vector-like lepton can decay by $E\to Z'\mu$, comparably to $E\to W\nu, Z\mu$.  In this case, for a small gauge kinetic mixing,  $Z'$ decays dominantly into $\mu{\bar\mu}$. As a result, there are six leptons from the decays of the vector-like lepton pair at the LHC. 
The vector-like lepton can also decay into $h\mu$ and $\varphi \mu$, but they are suppressed at the level of ${\rm BR} \lesssim 10^{-6}$, as shown in the right plot of Fig.~\ref{fig:VL-LHC}.
On the other hand, for $m_{Z'}>M_E/2$, the vector-like lepton decays dominantly by $E\to W\nu, Z\mu$.  In this case, from the leptonic decays of $Z$-boson, there are multi-leptons from the decays of the vector-like lepton pair at the LHC. 

Therefore, for both light and heavy $Z'$, the limits from the multi-lepton searches at LHC \cite{multilepton} apply. But, for the case with light $Z'$ having a larger BR with multi-leptons due to  $E\to Z'\mu$, the limit on the vector-like lepton mass is stronger. The ATLAS analysis for supersymmetry particles has been recently recast to set the limit on the vector-like lepton mass to $M_E>1\,{\rm TeV}$ \cite{kawamura2} for an $SU(2)_L$ singlet vector-like lepton.
Even if such a strong limit on the vector-like lepton mass is imposed, we can still explain the deviation of the muon $g-2$ due to the $Z'$ interactions, thanks to the non-decoupling phenomenon for the vector-like lepton, as discussed in the previous section.

Before closing the section, we also remark new decay modes of extra scalars due to the vector-like lepton in our model as compared to the case in Type-X 2HDM.
If $M_E<m_{h_i}, m_{H^-}$, with $h_i=H, A,\varphi$, the extra scalars can decay leptonically by $h_i\to E {\bar l}, {\bar E}l, l{\bar l}, E{\bar E}$ and $H^-\to E\,{\bar\nu}, l\,{\bar \nu}$, on top of the non-leptonic decay modes in Type-X 2HDM, $h_i\to q{\bar q}, WW, ZZ, W^\pm H^\mp$ for $h_i=H, A$. Thus, there are multi-lepton signatures from the decay of the extra scalars, which could be used to test our model with the vector-like lepton.

\section{Conclusions}

We have discussed the roles of the vector-like lepton for the generation of the muon mass through the seesaw mechanism in lepton-specific 2HDMs with a local $U(1)'$ symmetry. In this scenario, we showed that the non-decoupling effects of the vector-like lepton can be used to explain the muon $g-2$ anomaly at one-loop due to the light gauge boson $Z'$ and the light dark Higgs boson $\varphi$. We also found that there are overall specific couplings of $Z'/\varphi$ and the vector-like lepton to the muon, due to the mixing between the muon and the vector-like lepton.
Since the mixing angles of the vector-like lepton are tied up with the muon mass and the vector-like lepton mass by the seesaw mechanism, we can test the scenario by the muonic channels in various ways.

We found that the electroweak precision and Higgs data constrain the vector-like lepton relatively weakly, but the collider bounds on the vector-like lepton can be significant due to multi-lepton signatures. Moreover, we showed that the bounds from lepton flavor violating decays strongly constrain the couplings of the other leptons to the vector-like lepton.
On the other hand, we can evade the current LHC and collider bounds on the light $Z'$ and dark Higgs in our model, because they couple dominantly to the muon through the mixing of the vector-like lepton for a small gauge kinetic mixing and a small Higgs mixing, respectively.  Therefore, in order to probe the parameter space with the light $Z'$ and dark Higgs for explaining the muon $g-2$ anomaly, it is important to look for the light resonances with muon channels together with the vector-like lepton at LHC and future collider experiments.

\section*{Acknowledgments}

The work is supported in part by Basic Science Research Program through the National Research Foundation of Korea (NRF) funded by the Ministry of Education, Science and Technology (NRF-2019R1A2C2003738). 
This work of KY is supported by Brain Pool program funded by the Ministry of Science and ICT through the National Research Foundation of Korea(NRF-2021H1D3A2A02038697).
The work of JS is supported by the Chung-Ang University Graduate Research Scholarship in 2020.




\end{document}